\documentclass[journal]{IEEEtran}

\usepackage{amsmath, amsfonts}
\usepackage{algorithmic}
\usepackage{algorithm}
\usepackage{array}
\usepackage{textcomp}
\usepackage{stfloats}
\usepackage{url}
\usepackage{verbatim}
\usepackage{graphicx}
\usepackage{cite}
\usepackage{hyperref}
\usepackage{subfigure}
\usepackage{xcolor}

\usepackage{newtxmath}
\allowdisplaybreaks
\newtheorem{theorem}{\textbf{Theorem}}

\newtheorem{remark}[theorem]{\textbf{Remark}}

\usepackage{pifont}
\newcommand{\cmark}{\ding{51}}%
\newcommand{\xmark}{\ding{55}}%

\begin{document}

\title{Vehicle-to-Grid Fleet Service Provision\\considering Nonlinear Battery Behaviors}

\author{ Joshua Jaworski, Ningkun Zheng,~\IEEEmembership{Student Member,~IEEE}, Matthias Preindl,~\IEEEmembership{Senior Member,~IEEE}, Bolun~Xu,~\IEEEmembership{Member,~IEEE}
	\thanks{J. Jaworski, N.~Zheng, M. Preindl and B.~Xu are with Columbia University, NY, USA. Contact: \{jjj2146, nz2343, mp3501, bx2177\}@columbia.edu.}
} 



\maketitle

\begin{abstract}
The surging adoption of electric vehicles (EV) calls for accurate and efficient approaches to coordinate with the power grid operation. By being responsive to distribution grid limits and time-varying electricity prices, EV charging stations can minimize their charging costs while aiding grid operation simultaneously. In this study, we investigate the economic benefit of vehicle-to-grid (V2G) using real-time price data from New York State and a real-world charging network dataset. We incorporate nonlinear battery models and price uncertainty into the V2G management design to provide a realistic estimation of cost savings from different V2G options. 
The proposed control method is computationally tractable when scaling up to real-world applications. We show that our proposed algorithm leads to an average of 35\% charging cost savings compared to uncontrolled charging when considering unidirectional charging, and bi-directional V2G enables additional 18$\%$ cost savings compared to unidirectional smart charging. Our result also shows the importance of using more accurate nonlinear battery models in V2G controllers and evaluating the cost of price uncertainties over V2G.
\end{abstract}


\section{Introduction}
The International Energy Agency's (IEA) roadmap to achieve net zero greenhouse gas (GHG) emissions by 2050 calls for an increase in the share of renewable energy in total global power generation and global transport sector electrification from 29\% and 2\% in 2020 to 90\% and 45\% in 2050, respectively~\cite{iea2021,Bilgin2015}. However, increasing energy supply intermittency and electric vehicles (EV) charging demand can put significant stress on the grid without adequate management and control, such as peak charging demand during periods of low wind and solar power generation~\cite{ipcc2022}. Smart charging integrates external data, such as distribution grid constraints or time-varying electricity prices, into unidirectional (V1G) or bidirectional (V2G) power transfer management between the grid and the EV charging station (EVCS)~\cite{nrelSC}. Smart charging and V2G management have emerged as a key strategy to accelerate transportation electrification to support an increasingly renewable-powered grid operation, minimizing EV owners' charging cost, and leading new business models and job opportunities~\cite{ipcc2022,nour2019,szinai2020,EVSC2022}.


As the cost of V2G-compatible chargers continues to decline~\cite{jahnes2022design}, software development becomes pivotal to efficiently aggregate EVs and optimally control their V2G responses while meeting the designated charging targets. While plenty of works have conducted techno-economic analyses (TEA) of V2G ~\cite{long2021ord, cao2022, mouli2017integrated, ebrahimi2020tte}, few have considered complicating factors that practical V2G implementations must address. We group these factors into three categories. The first is battery model nonlinearities, in which the battery voltage, current, efficiency, and degradation depend on the state of charge (SoC). Controlling EV batteries accurately according to their nonlinear characteristics is crucial to strike a balance between ensuring battery security and economic benefits~\cite{GONZALEZCASTELLANOS2020}. The second is grid uncertainty, that the distribution grid load and electricity prices are time-varying and uncertain~\cite{EVSC2022}. Uncertainties are often neglected in TEA primarily due to computation difficulties, but practical V2G implementations must consider uncertainties in price-response applications. The last is computational scalability, the V2G management software must manage tens to hundreds of EVs without consuming monstrous computing power. As we will later be shown in our results, while the aggregate benefit of V2G is pivotal for future grids, economic saving for each individual EV is not significant to justify investment in specialized computing hardware. 

This paper presents a computation-efficient V2G management framework and a realistic case study integrating the aforementioned complicating factors in practical V2G implementations. Our contributions include:
\begin{itemize}
    \item We propose a computation-efficient and scalable V2G management controller which optimizes V2G charging using accurate nonlinear battery models under stochastic electricity prices. Combining a stochastic dynamic programming algorithm with a least-laxity first (LLF) scheduling algorithm~\cite{llf2017}, our proposed V2G framework minimizes charging costs for EVCS to meet charging targets and distribution grid limits.
    \item Using real-world electricity price and EV charging behavior data, our paper provides a first-of-its-kind case study to demonstrate  cost savings an EVCS can realistically achieve in various V2G settings.
    \item Our case study compares uncontrolled charging, V1G, and V2G with and without nonlinear storage models and price uncertainties. The results quantify the impact of various charging and model options and guide EVCS planning and technology developments.
\end{itemize}
The remainder of this paper is organized as follows. Section~\ref{sec:lr} presents the literature review. Section~\ref{sec:form} describes the system model and formulates the EV charging cost minimization problem. Section~\ref{sec:solu} presents the solution algorithm to the formulated charging problem. Section~\ref{sec:cs} includes simulation results and discussion. Section~\ref{sec:conc} concludes the paper.

\section{Literature Review}\label{sec:lr}
Previous literature has proposed multiple heuristic, optimization, or learning-based approaches to conduct smart charging mostly considering  linear battery models with constant power rating and efficiency~\cite{FACHRIZAL2020, abdullahRL2021}. For example, Liu et al.~\cite{liu2021tte} formulate an EVCS controller as a bi-level program and use a genetic algorithm to minimize charging costs under a time-of-use (ToU) tariff. Similarly, Long et al.~\cite{long2021ord} use an ordinal optimization approach to minimize EVCS operating cost under a ToU  scheme but add aggregated EV demand, maintenance costs, V2G capability, hydrogen storage, and renewable energy generation to the formulation.  Additionally, Cao et al.~\cite{cao2022} propose a custom actor-critic algorithm to minimize charging costs and peak charging load for a V2G-enabled EVCS, which results in a 24\% energy cost savings when compared to uncontrolled charging.

While most prior V2G literature included benchmarks to demonstrate the effectiveness of the proposed algorithm~\cite{liu2021tte, long2021ord, cao2022}, they often assume constant EV battery parameters (power ratings, efficiency) and do not include a penalty term to minimize battery cycling. Lab experiments and real-world data have shown that Li-ion battery power ratings and efficiency strongly depend on SoC, especially in nickel-cobalt-based batteries, which are the most common choice for  EVs~\cite{wuwang2020,zhang2018, preger2020degradation}. In EV smart charging or V2G applications, which aim to provide high charging power with low-cost power conversion hardware. Battery power rating and efficiency are sensitive to the SoC, and the charge or discharge power must be carefully controlled to ensure battery thermal security and reduce degradation rates~\cite{zhou2022robust}. A common protocol for EV charging management is the CC--CV (constant current--constant voltage) method~\cite{boraCCCV2016,habibCCCV2018,Eull2021}, that the battery charges with constant current until reaching a high SoC level and then gradually reduces the current to maintain a constant charging voltage to prevent over-voltage damages. 
Modeling battery characteristics such as CC--CV protocols in V2G management is critical to maximize cost savings and ensure battery security, but it requires representing battery power and efficiencies as functions of SoC instead of using constant values, which introduces significant computation complexities~\cite{pandzic2019, FARAG2017, SAKTI2017, rancilio2019, GONZALEZCASTELLANOS2020}.

Some EVCS smart charging algorithms partially accounted for nonlinear charging/discharging characteristics~\cite{mouli2017integrated ,ebrahimi2020tte, schwenk2021, lee2021adaptive}, but few were able to model all nonlinear factors in a computation efficient approach. Starting from an EVCS profit maximization problem formulated as mixed-integer linear programming (MILP), Mouli et al.~\cite{mouli2017integrated} accounts for SoC-dependent power ratings in which the maximum power drops linearly  after 80\% SoC. Ebrahimi et al.~\cite{ebrahimi2020tte} models battery degradation as dependent on both SoC and depth of discharge (DoD), and Schwenk et al.~\cite{schwenk2021} incorporated both nonlinear efficiency and nonlinear degradation terms. Lee et al.~\cite{lee2021adaptive} implemented an adaptive scheduling algorithm that formulates custom objectives and constraints as a convex program and computes an optimal charging schedule in real-time while considering feeder limits and battery tail capacity reclamation at high SoCs with a data-driven approach. However, the reviewed control solutions do not model all power ratings, efficiencies, and cycling penalties/degradation as nonlinear nor do they provide a method to incorporate different behavior curves. This property will be pivotal for the adaptability of EVCS control algorithms to fast-charging applications and manufacturer-customized EV battery management systems (BMS).

Besides battery models, price uncertainty is another complicating factor that may impact the EVCS cost estimations but was rarely studied in V2G due to computation difficulties.
Most literature on smart charging assumes perfect price forecasts or predetermined ToU tariffs~\cite{liu2021tte, long2021ord, cao2022, mouli2017integrated ,ebrahimi2020tte, schwenk2021, lee2021adaptive}. As EV capacity surges, future V2G projects will most likely arbitrage real-time electricity prices that are highly volatile and uncertain, and EVCS must consider price uncertainties.
Frendo et al. and Ahmad et al. ~\cite{frendo2019, ahmad2019tte} use day-ahead prices to forecast the next day's LMPs, incorporating EVCS controller using MILP formulation. Zhang et al.~\cite{zhang2021} develop the charging control deep deterministic policy gradient, which models EV charging as a Markov decision process (MDP) and optimizes user satisfaction and charging costs with the output of a long short-term memory (LSTM) network that approximates sequential energy price dynamics. Results from this research demonstrated the significance of modeling price uncertainties, but the computing approach is not scalable to address nonlinear storage models.

As summarized in Table \ref{tab:litreview}, the reviewed literature proposes EVCS algorithms that partially account for nonlinear EV battery behavior and price uncertainty. We close this research gap by developing a real-time EVCS V2G control algorithm based on analytical nonlinear stochastic dynamic programming (SDP) and a least-laxity first (LLF) scheduling approach that adds nonlinear EV battery behavior and price uncertainty to the EVCS control formulation while minimizing operating costs and complying with EV charging and battery dynamics such as CC-CV charging profiles, facility power limits, and users' charging targets. While our proposed method accounts for system non-linearity and uncertainty, we demonstrate that it is scalable and computationally tractable in practical scenarios. 

\begin{table}[t]
\footnotesize
\caption{Consideration of price uncertainty and battery parameters nonlinear behavior in smart charging control formulations}
\centering
\begin{tabular}{lcccc}
\hline
\hline
\shortstack{} & \shortstack{Price\\Uncertainty} & \shortstack{Nonlinear\\Power\\Rating} & \shortstack{Nonlinear\\Efficiency} & \shortstack{Nonlinear\\ Cycling Penalty / \\Degradation Cost}\\
\hline
\cite{liu2021tte, long2021ord, cao2022}  &\xmark   &\xmark   &\xmark   &\xmark\\
\cite{mouli2017integrated, lee2021adaptive}   &\xmark   &\cmark   &\xmark   &\xmark\\
\cite{ebrahimi2020tte}      &\xmark   &\xmark   &\xmark   &\cmark\\
\cite{schwenk2021}      &\xmark   &\xmark   &\cmark   &\cmark\\
\cite{frendo2019, ahmad2019tte, zhang2021}     &\cmark   &\xmark   &\xmark   &\xmark\\
Proposed    &\cmark   &\cmark   &\cmark   &\cmark\\
\hline
\hline
\end{tabular}
\label{tab:litreview}
\end{table}

\section{System Model and Formulation}\label{sec:form}
We take the perspective of a public EVCS operator whose electricity cost settles using  time-varying wholesale electricity real-time prices. The objective of the EVCS is to minimize the electricity cost under either smart charging (V1G) or V2G operation. The EVCS has enough chargers, all with the same specifications so that no rejection-of-service event occurs during the simulation time frame and complies with $L$, the maximum power rating of the charging station, at all time steps.

\subsection{EV Charging Sessions}
We consider a total of $K$ EVs accessing the EVCS during the considered period, $\mathcal{K}=\{1,..,K\}$ is the set of EVs. EV $k\in\mathcal{K}$ arrives on time step $A_k$ with a starting SoC of $S_k$, and departs on time $D_k$ with a charging target SoC $F_k$, with  $D_k > A_k$ and $F_k > S_k$. Thus we use $\mathcal{T}_k = \{A_k,\dotsc,D_k\}$ to denote the time frame of the current charging session of EV $k$. At every EV arrival, the controller updates the tuple ($A_k$, $D_k$, $F_k$), which is used as input for the proposed solution algorithm. Additionally, the controller has access to every EV SoC at all times. We also assume the EVCS does not have information or can predict the arrival of EVs, but each EV upon arrival will inform the EVCS its departure time and charging target.

\subsection{Battery Nonlinear Behavior}
\begin{figure}[ht]
\centering
\includegraphics[width=1\columnwidth]{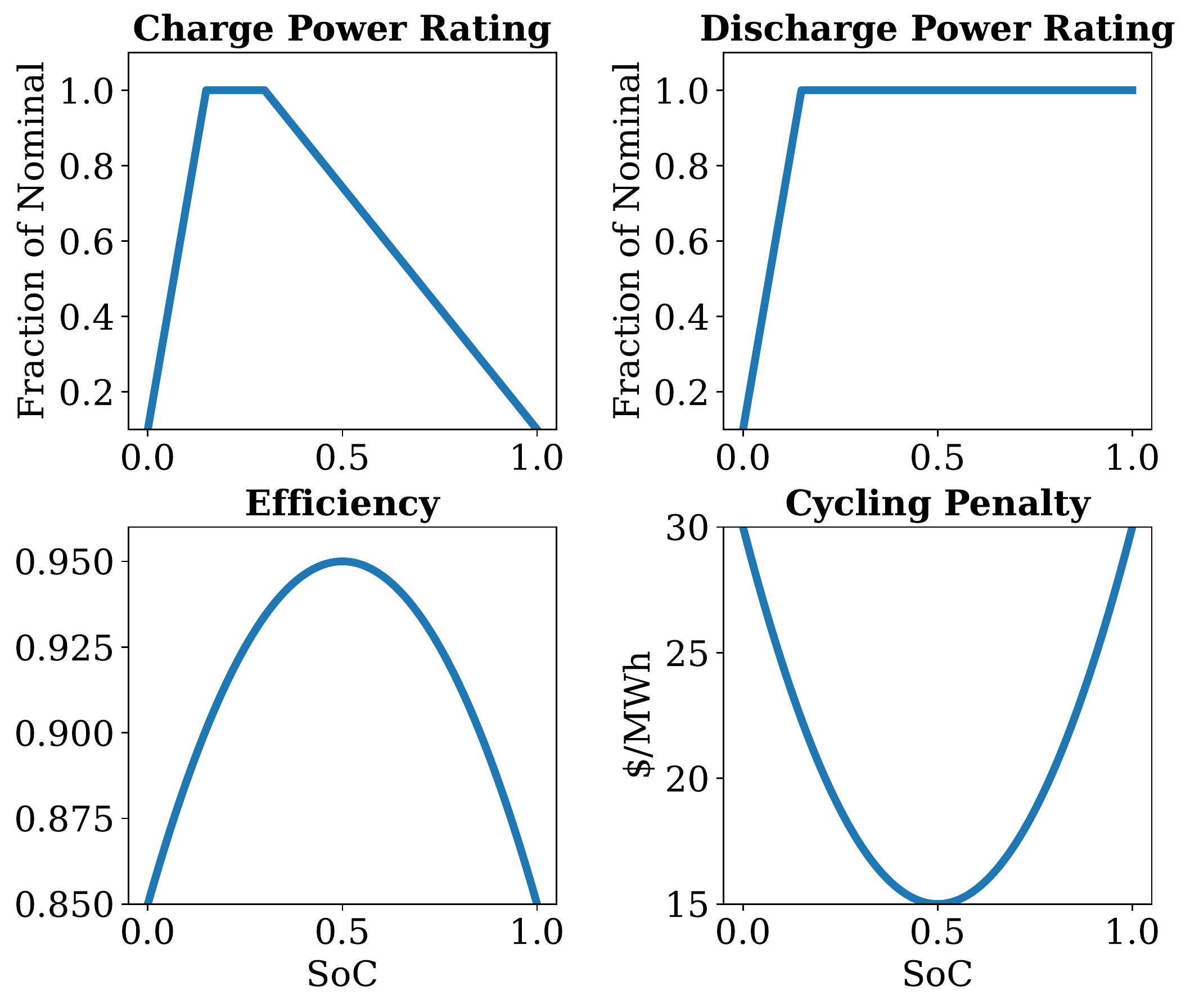}
\caption{Assumed EV battery SoC-dependent parameter curves for all EVs}
\label{Fig.Curves}
\end{figure}

All EVs are modeled as having the same battery capacity and SoC-dependent charge power rating ($B_k(e_{t-1,k})$), discharge power rating ($P_k(e_{t-1,k})$), single-trip efficiency ($\eta_k (e_{t-1,k})$) and discharge cost penalty ($c_k(e_{t-1,k})$) curves. As shown in Figure~\ref{Fig.Curves}, efficiency, and cycling penalty are modeled as quadratic functions of SoC~\cite{wuwang2020}. The power rating curves resemble a Tesla Model S fast charging curve~\cite{teslacurve21} with custom-defined CC-CV behavior at low and high SoC. 
The controller has access to a low-resolution version of the battery nonlinear parameter curves to use as an approximation during control policy computations. This resolution gap simulates the challenge of EV battery behavior approximation in online smart charging control.


\subsection{Price Prediction}\label{subsec:price}
The EVCS operator can access the system's day-ahead price (DAP) information and a DAP-based real-time price (RTP) prediction tool. For this work, RTP is modeled as a 1st-order Markov process with 12 nodes per time step trained with historical RTP-DAP bias price data as in setting DB-Dep in~\cite{zjx2022}. The proposed solution algorithm uses the resulting prediction to provide a control decision that accounts for future price uncertainty.

\subsection{Formulation}
We start with formulating the EVCS V2G fleet management problem including nonlinear battery models, charging station power limits, and a causality control policy constraint with respect to time-varying electricity price uncertainties.
\begin{subequations}\label{eq2}
The objective of the charging station is to minimize the cost of electricity to charge up each EV including a discharge penalty to avoid frequent cycles that accelerate battery degradation
\begin{align}
    \min_{b_{t,k},p_{t,k}} \sum_{k\in \mathcal{K}} \sum_{t\in \mathcal{T}_k} \lambda_t (b_{t,k}-p_{t,k}) + c_{k}(e_{t-1,k}) p_{t,k}
\end{align}
in which $b_{t,k}$ is the charging power of EV $k$ during time step $t$, while $p_{t,k}$ is the discharge power. $\lambda_t$ is the time-varying price of electricity. The second term reduces battery degradation by introducing a discharge penalty cost $c_{k}(e_{t-1,k})$ as a function of SoC. While the EVCS operator does not assume EV battery degradation cost, the controller incorporates an EV battery discharge penalty to avoid excessive cycling. Note that $b_{t,k}$ and $p_{t,k}$ have been normalized by the chosen simulation time step and have units of energy.

Each EV is subject to the following power and energy constraints ($t\in \mathcal{T}_k$)
\begin{align}
    0 \leq b_{t,k} \leq B_k(e_{t-1,k}) \label{p1_c1}\\
    0\leq p_{t,k} \leq P_k(e_{t-1,k}) \label{p1_c2} \\
    e_{t,k} - e_{t-1,k} = -p_{t,k}/\eta_k (e_{t-1,k}) + b_{t,k}\cdot \eta_k (e_{t-1,k}) \label{p1_c3}\\
    \underline{E}_k \leq e_{t,k} \leq \overline{E}_k \label{p1_c4}\\
    e_{A_k} = S_k,\quad e_{D_k} \geq F_k \label{p1_c5}    
\end{align}
in which $e_{t,k}$ is the SoC of EV $k$ during time step $t$. $\underline{E}_k$ is the EV minimal SoC limit and $\overline{E}_k$ is the EV maximum SoC. \eqref{p1_c1} and \eqref{p1_c2} model the EV power rating, \eqref{p1_c3} models the SoC evolution, \eqref{p1_c4} models the upper and lower SoC limits, and \eqref{p1_c5} models the starting SoC and SoC charging target.

The total charging and discharge power is subjected to the station power limit
\begin{align}
    \sum_{k\in \mathcal{K}} p_{t,k} &\leq L \label{p1_c6}\\
    \sum_{k\in \mathcal{K}} b_{t,k} &\leq L \label{p1_c7}
\end{align}

The control policy must be causal (non-anticipatory)~\cite{shapiro2021lectures} and only depends on past and current information
\begin{align}
\{b_{t,k},p_{t,k} | k\in \mathcal{K}, t\in \mathcal{T}_k\} \in \text{Causal Control Policies}
\end{align}
\end{subequations}

\begin{remark}\textbf{Generalization to different charging scenarios.}
\eqref{eq2} provides a generalized formulation in different charging scenarios. In V2G, both $B_k$ and $P_k$  are non-zero, while in single-directional smart charging, or V1G, the EV will not inject power into the grid and thus $P_k$ is set to zero. In cases when assuming a linear battery model, $P_{k}$, $B_{k}$, $c_{k}$ and $\eta_k$ are constants, while in nonlinear battery models, these parameters are dependent on the SoC.
\end{remark}

\section{Solution Method}\label{sec:solu}
We take a two-step approach to solving the EVCS control problem~\eqref{eq2}. First, we formulate the charging session of a single EV as a price arbitrage problem, which satisfies all constraints in \eqref{eq2} except the EVCS power limit constraints \eqref{p1_c5} and \eqref{p1_c6}, and solve it using an analytical SDP method ~\cite{zjx2022}. Second, we aggregate the control policies from the EVs in an active session at the current time step and prioritize the distribution of control signals according to an LLF approach ~\cite{llf2017} to ensure compliance with the EVCS power limit.

\subsection{Decomposition to Arbitrage Problems}
The core of the proposed control method is a single energy storage price arbitrage problem formulated as an SDP~\cite{zjx2022}.
\begin{remark}\textbf{V2G decomposition.}
Because we do not assume the EV charging actions would impact market clearing prices, we can decompose \eqref{eq2} into parallel arbitrage problems by relaxing the charging station power limit constraint \eqref{p1_c5} and \eqref{p1_c6}, which are the only coupling factors among all EVs. Each resulting sub-problem becomes an arbitrage problem that maximizes the arbitrage profit (or, equivalently, minimizes the charging session electricity cost) while meeting the final SoC target. We will discuss in later sections how we aggregate results from all EVs to incorporate the EVCS power limits.
\end{remark}

To solve the arbitrage sub-problem under the causality policy constraint, we adopt an SDP approach with the following formulation (for simplicity, we omit the EV index $k$ in the following formulation, but the following formulation is for a single EV):
\begin{subequations}\label{eq1}
\begin{align}
    Q_{t-1}(e_{t-1}\,|\,\lambda_t) &= \max_{b_t, p_t} \lambda_t (p_t-b_t) - c_k(e_{t-1})p_t + V_t(e_t\,|\,\lambda_t) \\
    V_t(e_t\,|\,\lambda_t) &= \mathbb{E}_{\lambda_{t+1}}\Big[Q_{t}(e_{t}\,|\,\lambda_{t+1}) \Big| \lambda_t\Big] \label{eq:obj4}
\end{align}
subject to \eqref{p1_c1}--\eqref{p1_c4}.
\end{subequations}
We model the time-varying price $\lambda_t$ as an order-1 Markov process, as described in \ref{subsec:price}, in which the price distribution over a time period $t+1$ depends on the realized price over $t$. $Q_{t-1}(e_{t-1}\,|\,\lambda_t)$ is the maximized current period profit given the SoC at the start of the time period, while  $V_t(e_t\,|\,\lambda_t)$ is the expected value function representing the opportunity value of energy stored in the battery at the end of the time period. In the context of an EV charging session, $V_t(e_t\,|\,\lambda_t)$ represents the minimum expected cost of the remainder of the session based on the current SoC, current RTP, and the RTP uncertainty model.


\subsection{Solving Nonlinear Battery Models}

We extend the solution approach from~\cite{zjx2022} by incorporating the SoC-dependency of battery behavior parameters into the first-order optimality condition expression as
\begin{align}\label{eq3}
&q_{t-1,i}(e_{t-1}) = \nonumber\\
    &\pi_{t,i} \Big(\frac{\partial p_t}{\partial e_{t-1}}-\frac{\partial b_t}{\partial e_{t-1}}\Big) - c_k(e_{t-1})\frac{\partial p_t}{\partial e_{t-1}}
    -\frac{\partial c_k(e_{t-1})}{\partial e_{t-1}}p_t
    \nonumber\\
    &+ v_{t,i}(e_t)\frac{\partial e_t}{\partial e_{t-1}} = 0
\end{align}
where $q_{t-1,i}(e_{t-1})$ is the derivative of $Q_{t-1}(e_{t-1}\,|\,\lambda_t)$, or the storage device's marginal opportunity value. According to the Karush-Kuhn-Tucker (KKT) conditions and \eqref{p1_c3}, we obtain the following:
\begin{subequations}\label{eq4}
\begin{align}
    &{\partial p_t}/{\partial e_{t-1}} = \\
    &\begin{cases}
    \eta+(p_t/\eta){\cdot}({\partial \eta}/{\partial e}) \indent\indent &\quad \text{if \eqref{p1_c2} not binding} \\
    {\partial P}/{\partial e} &\quad \text{if \eqref{p1_c2} binding} \nonumber
    \end{cases}\\
    &{\partial b_t}/{\partial e_{t-1}} = \\
    &\begin{cases}
    -1/\eta-(b_t/\eta){\cdot}({\partial \eta}/{\partial e}) \indent\indent &\text{if \eqref{p1_c1} not binding} \\
    {\partial B}/{\partial e} &\text{if \eqref{p1_c1} binding} \nonumber
    \end{cases}\\
    &{\partial e_t}/{\partial e_{t-1}} = \\
    &\begin{cases}
    0 \qquad\qquad\qquad\qquad\qquad\qquad \text{if \eqref{p1_c1} or \eqref{p1_c2} not binding} \\
    1-(1/\eta){\cdot}({\partial P}/{\partial e})+(1/\eta)^2{\cdot}({\partial \eta}/{\partial e})P \\ + \eta{\cdot}({\partial B}/{\partial e}) + B{\cdot}{\partial \eta}/{\partial e} \qquad\quad \text{if \eqref{p1_c1} and \eqref{p1_c2} binding} \nonumber
    \end{cases}
\end{align}
\end{subequations}

By replacing the partial derivative expressions given by \eqref{eq4} in \eqref{eq3} for full power rating (binding) and partial (non-binding) charging or discharging cases, we obtain an analytical marginal opportunity value function update expression. Note that the expressions in \eqref{eq4} involve the optimization variables $p_{t}$ and $b_{t}$. The expressions are solved by approximating $p_{t}$ and $b_{t}$ by the power ratings $B$ and $P$ corresponding to the current SoC. The full formulation of this equation is deferred to Appendix~\ref{app:vf}.

\subsection{Arbitrage Policy}\label{CP}
We use the developed analytical SDP algorithm to calculate a marginal opportunity value function for each charging session for each EV. At each time step, the control decision $p_{t}$ and $b_{t}$ for each connected EV can be determined by comparing the corresponding marginal opportunity value function and the observed realized RTP $\lambda_t$. The value difference between the EV battery's marginal value and the marginal grid price will trigger a charging, discharging, or idling control signal. The full marginal value function and control policy calculation methods are deferred to Appendix~\ref{app:cp} and Appendix~\ref{app:vf}.

\begin{remark}\textbf{Lagrangian relaxation of the final SoC constraint.}
We apply a Lagrangian relaxation to incorporate the final SoC charging target constraint into the SDP by assuming an arbitrarily large penalty (\$1000/MWh) for not achieving the charging target. This enables the marginal value function corresponding to the EV departing time step to act as an inverse activation function, with a linear penalty cost to the battery SoC until reaching the specified charging target. This enforces the battery to charge regardless of the price when approaching the end of the charging session in V1G and V2G cases to meet the charging target.
\end{remark}

When applying this control policy with a nonlinear system assumption is that the storage device parameters $B$, $P$, $\eta$, and c used in the elaboration of the control decision are approximations of the real storage device behavior parameters. The control decision will be calculated using a trained model from historical price data and executed in the testing environment. The testing environment will limit the control inputs to a range within the true storage behavior constraints. 

\subsection{EV Fleet Control Simulation}
The EVCS control algorithm incorporates nonlinear battery parameter curve approximations, data provided by the EV users (SoC charging targets and session duration), the marginal value function calculation algorithm, an LLF prioritization step, and the control policy outlined in \ref{CP}. The algorithm is executed in real-time as follows:
\begin{enumerate}
    \item Set $t\to t+1$.
    \item Calculate the marginal value function for the EVs that arrive to the EVCS at time $t$. For the value function computation, set $t$ as the starting time, the provided session duration as the time horizon $T$ and the provided SoC session target as the final SoC $e^{f}$. The resulting marginal value function will be used as the basis of the control policy for its corresponding EV for the duration of the current session.
    \item Identify all connected EVs. Compute the ratio of time elapsed in the current session to the total session duration for each connected EV. Based on the calculated ratios, sort the EVs in descending order. This step is aligned with an LLF~\cite{llf2017} scheduling approach, which prioritizes EVs with the least time to achieve their target.
    \item In the order defined by step 3), execute the control policy for each EV as described in \ref{CP} (i.e., comparing the EV marginal value to the grid marginal price). If the facility power limit is reached, set the remaining EVs' power control signals to zero and go to step 1).
    \item Go to step 1) until reaching the target simulation time.
\end{enumerate}

Note that the control signals in step 4) will be truncated by the testing environment if they are outside the range of the actual battery behavior model. The LLF sorting step provides a lightweight solution to aggregate the individual value function results and comply with the facility's power limit. This enables the modular nature of the algorithm components and prevents exponential computation time growth as the number of EVs in the EVCS increase.

\section{Case Study}\label{sec:cs}
\subsection{Data and Experiment Design}
We test the proposed control algorithm using the 2019 New York Independent System Operator (NYISO) price data. Price uncertainty is modeled using a 1st-order Markov process trained with 2016-2018 NYISO price data. We include prices from four zones to demonstrate performance results in different price patterns: NYC, LONGIL, NORTH, and WEST.


A 101-sample resolution version of the SoC-dependent battery parameter curves shown in Figure~\ref{Fig.Curves} is considered the ground truth and used as the testing environment. We assume that the controller has access to a 10-sample resolution version of the same battery nonlinear parameter curves. Although the proposed method can handle different parameter curves for each EV, we assume identical parameter curves in this case study for simplicity. With this resolution gap between the environment and the valuation process, we demonstrate the effectiveness of the proposed algorithm in providing efficient control with a limited amount of data to approximate the battery model.

We consider six scenarios to test the proposed algorithm and establish benchmarks for comparison. The following two scenarios are used as benchmarks:
\begin{enumerate}
    \item \textbf{PF (perfect forecast).} We perform a deterministic optimization using real-time prices in Julia/Gurobi. The optimization problem setup can be found in Appendix~\ref{app:milp}. This benchmark scenario represents the lowest possible EVCS operating cost.
    \item \textbf{UC (uncontrolled charging).} EVs start charging as soon as they arrive at the EVCS and charge with the maximum allowable power rating at all times until reaching their charging target. Facility power limits are fairly distributed among actively charging EVs. The control logic is implemented in Julia. This is the second benchmark case and represents an EVCS without a control policy or V2G capability. 
\end{enumerate}
and the remaining scenarios are solved with the proposed custom algorithm implemented in Julia:
\begin{enumerate}
    \item \textbf{NL-V2G.} We perform SDP control assuming V2G capability and approximate EV battery and charger behavior with a 10-sample version of the nonlinear parameter curves.
    \item \textbf{NL-V1G.} Similar to NL-V2G, but assuming no V2G capability. 
    \item \textbf{L-V2G.} Similar to NL-V2G, but approximating EV battery and nonlinear charger parameters with constant values. Power ratings are set to the nominal charger capability, one-way charging and discharging efficiencies to 95\% and marginal battery degradation cost to \$15\/MWh
    \item \textbf{L-V1G.} Similar to L-V2G, but assuming no V2G capability. 
\end{enumerate}

\begin{figure}[t]
\centering
\includegraphics[width= .8\columnwidth]{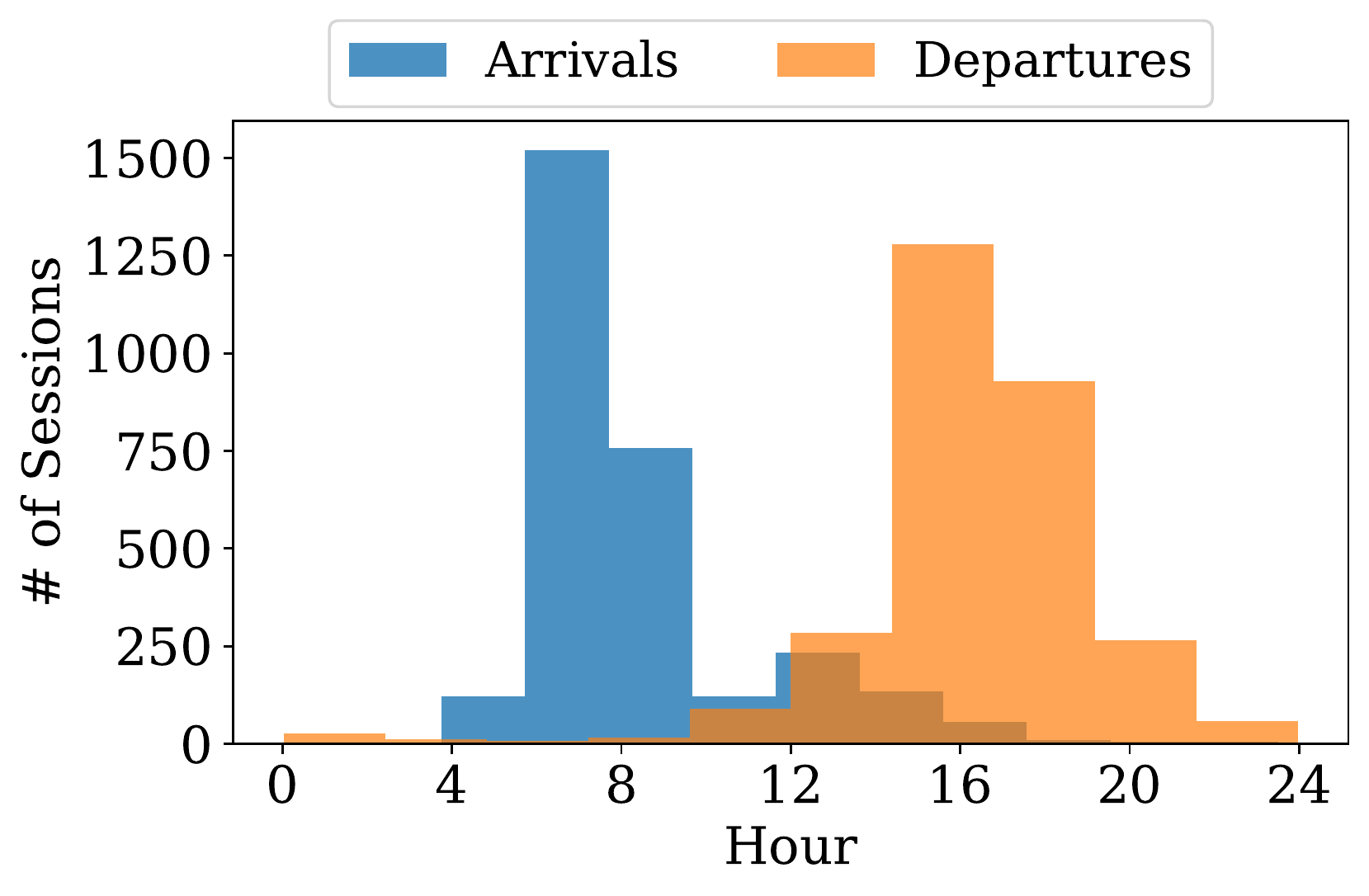}
\caption{Histogram of EVCS users' arrival and departure times from the ACN (JPL 2019) Dataset.}
\label{Fig.Arrivals}
\end{figure}

The EVCS consists of 21 bi-directional (unless otherwise noted by the scenario being tested) 17.2 kW level-2 chargers and has a 150 kW power limit, leading to an over-subscription ratio of 2.4. We assume 75 users have access to the EVCS and all users own a 100 kWh EV. Users' energy requested, and arrival and departure times are obtained from the Caltech ACN dataset~\cite{acndata2019}, specifically using the 2019 JPL data with energy requests greater than 5 kWh. Figure~\ref{Fig.Arrivals} shows the distribution of arrival and departure times of the dataset that will be used for the simulation. A starting SoC of 10\% is assumed for all arrivals. SDP control is performed in all scenarios using the described 1st-order Markov process price prediction.

\begin{remark}\textbf{Charging target compliance.} For a charging session to be successful in any of the scenarios, the control algorithm must achieve a final SoC within 5\% of the user's SoC charging target. We define the charging target compliance performance metric as the ratio between successful and total charging sessions. Note that the user may input a target that is infeasible due to a short session duration. The charging target compliance metric will exclude these infeasible cases.
\end{remark}

All computations were performed on a personal laptop with an Intel Core i9-10885H 2.5GHz CPU and 32 GB memory. The benchmark (PF) using MILP is solved using Gurobi~\cite{gurobi}, while the proposed algorithms and the EVCS simulation are implemented in Julia.

\subsection{Cost Savings and Charging Target Compliance}

\begin{figure}[t]
\centering
\subfigure[V2G]{\includegraphics[width=0.8\columnwidth]{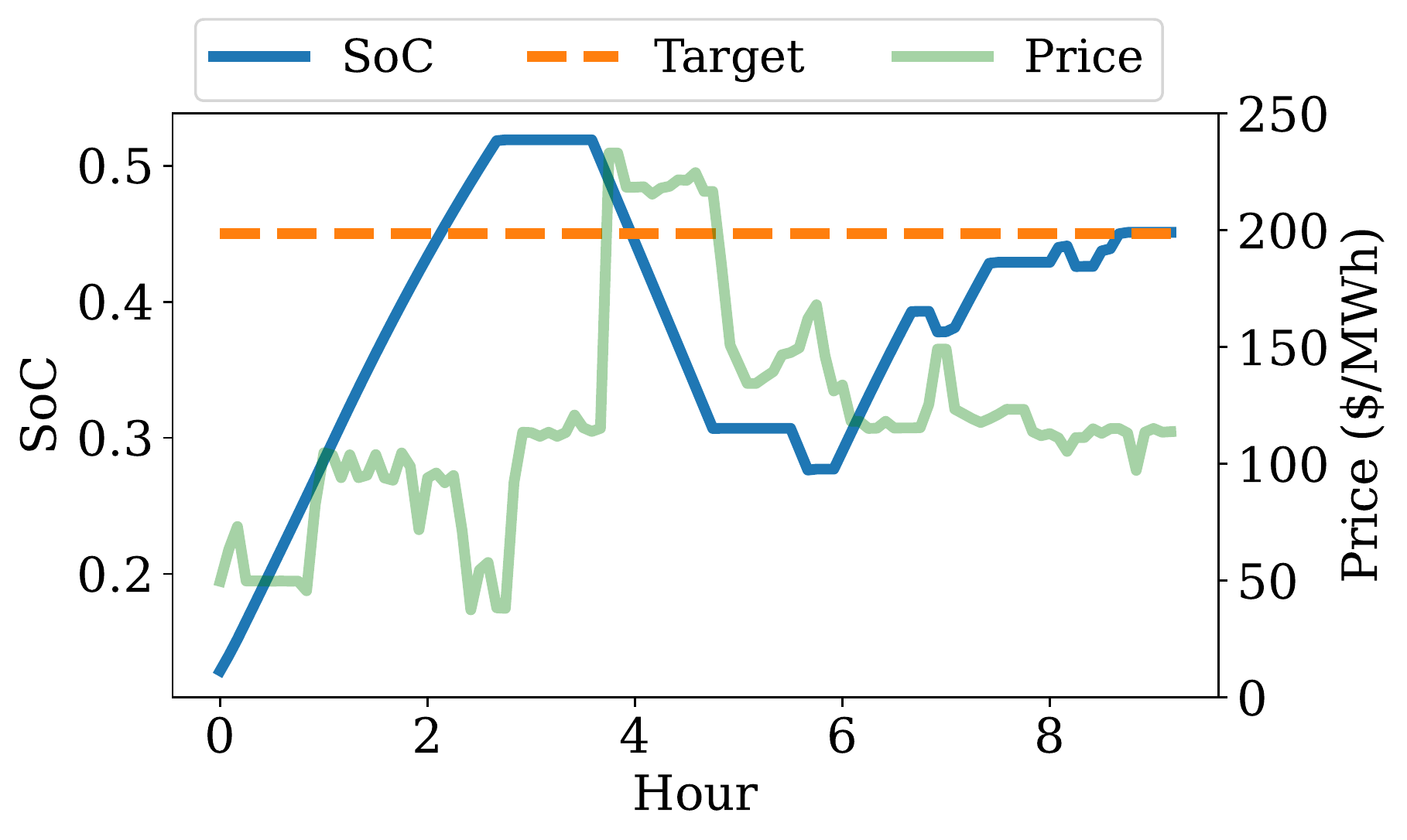}}
\subfigure[V1G]{
\includegraphics[width=0.8\columnwidth]{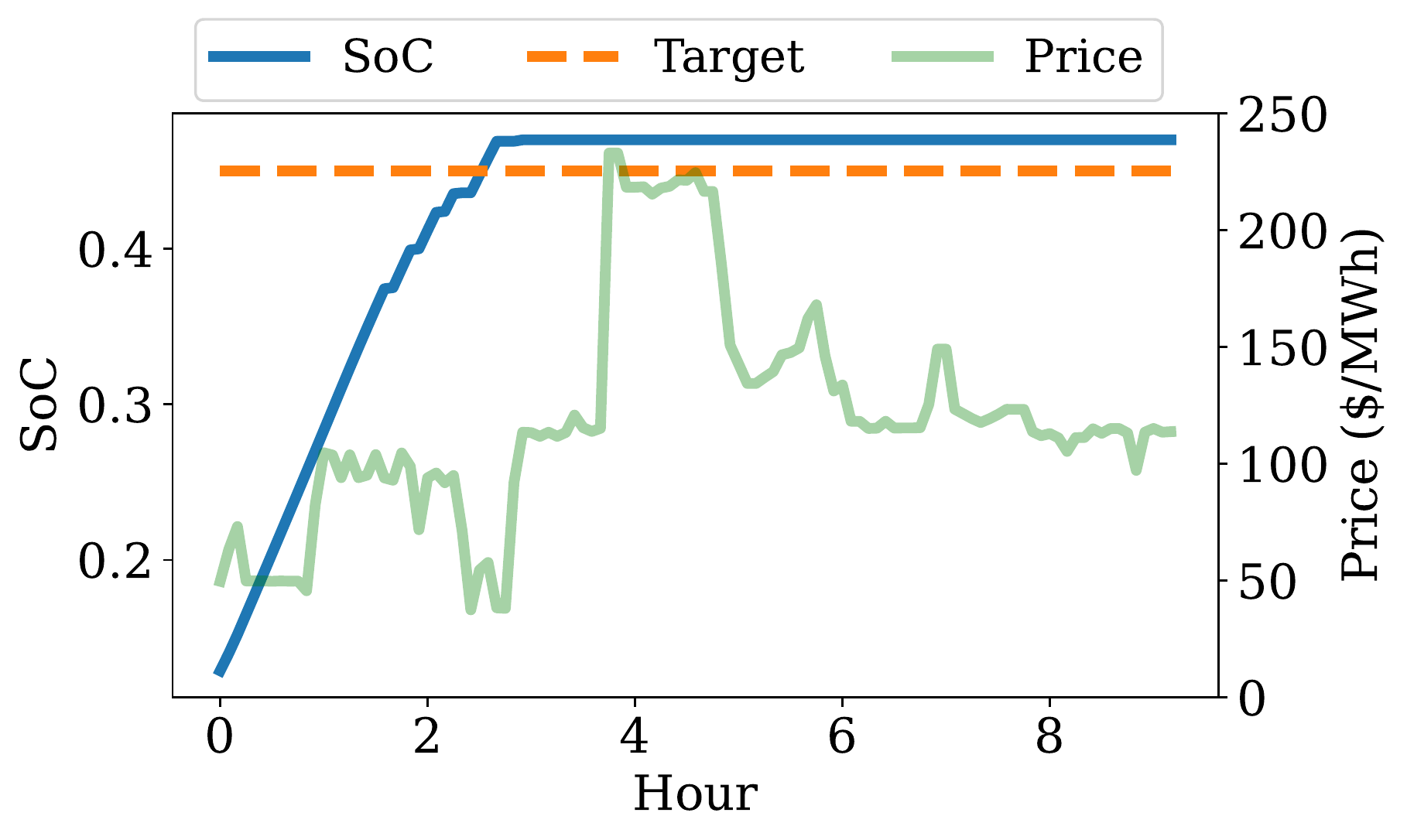}}
\caption{Performance of the proposed control policy in a 9-hour charging session in both (a) NL-V2G and (b) NL-V1G scenarios}
\label{Fig.SingleSession}
\end{figure}

Figure~\ref{Fig.SingleSession} shows a sample charging session using both NL-V2G and NL-V1G scenarios. V2G achieves cost reduction by charging during low price periods and capturing additional revenue through energy arbitrage, and V1G reduces EVCS operating costs only through its smart charging capability. Note that embedding the final SoC requirement in the value-to-go function calculation results in successful charging sessions for all the shown cases while minimizing total cost and EV battery cycling during the session.


\begin{figure*}[t]
    \centering
    \includegraphics[width=1.4\columnwidth]{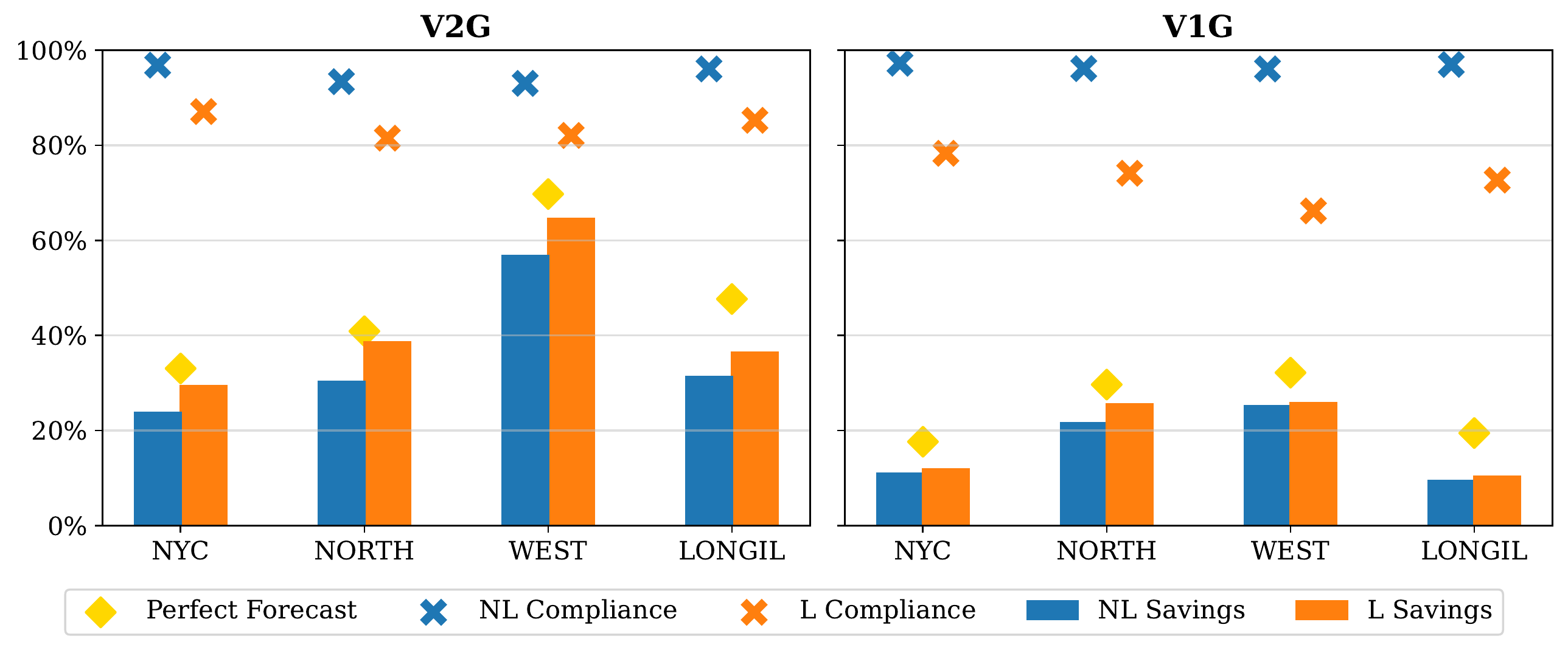}
    \caption{EVCS operating cost savings of all scenarios simulated with the proposed method for the uncontrolled case. Perfect Forecast represents the highest possible EVCS operating cost savings under the considered simulation conditions.}
    \label{Fig.CostResults}
\end{figure*}

\begin{figure*}[t]
    \centering
    \includegraphics[width=1.4\columnwidth]{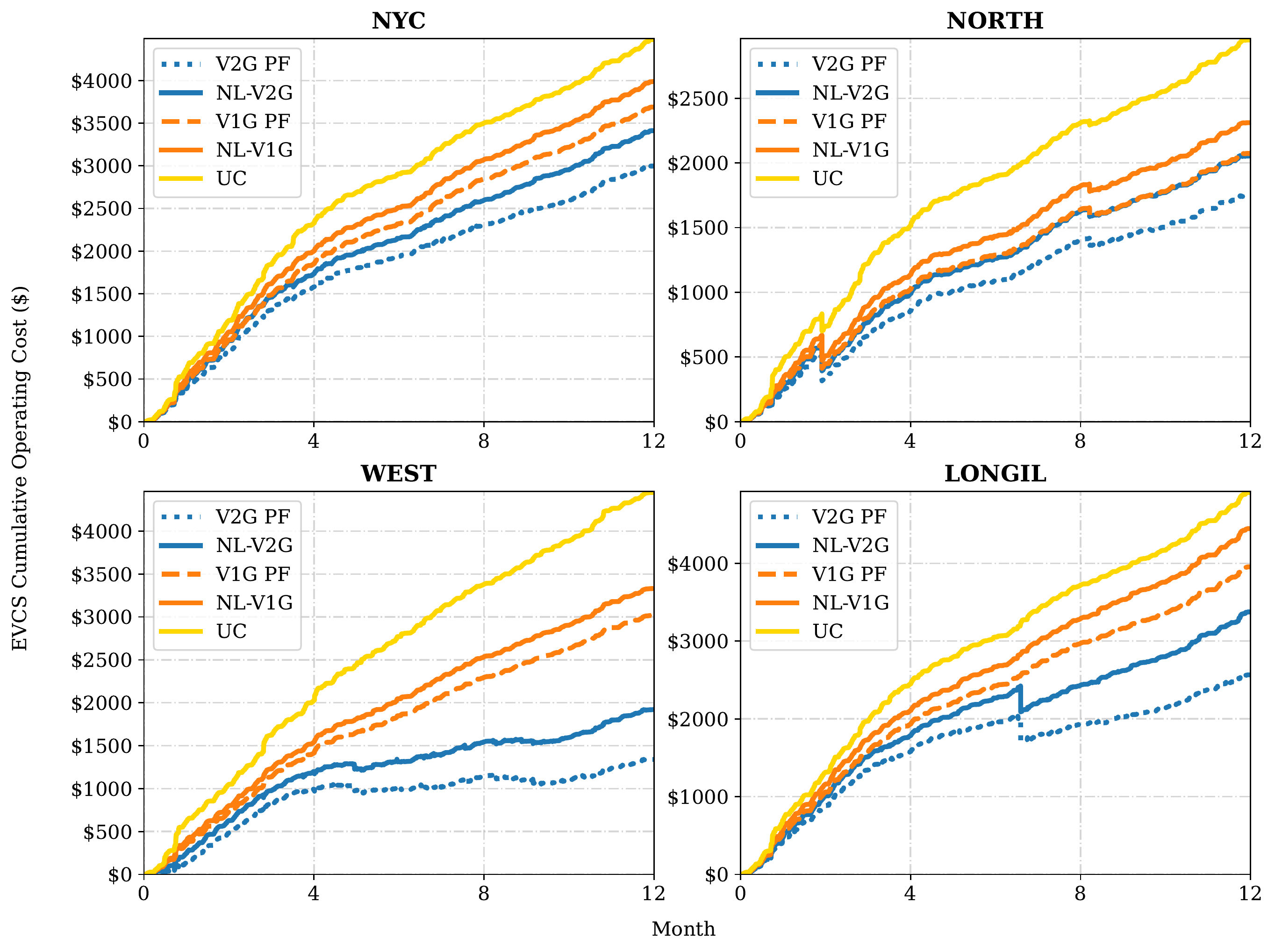}
    \caption{EVCS cumulative operating costs during 2019 for all simulated NYISO zones. Perfect-forecast V2G/V1G, stochastic V2G/V1G and uncontrolled charging scenarios are shown.}
    \label{Fig.CostCumulative}
\end{figure*}

Figure~\ref{Fig.CostResults} shows the EVCS operating cost savings achieved by the proposed algorithm and the PF scenario as a percentage of the uncontrolled charging (UC) scenarios as well as the charging compliance results across the four considered NYISO zones. NL-V2G results in an average operating cost savings of 35\% over UC, with savings reaching up to 56\% in the WEST zone while maintaining average charging compliance of 95\%. The cost savings average drops to 17\% across zones if NL-V1G is used. This demonstrates the impact of bi-directional charging capability on EVCS operating costs. Both L-V2G and L-V1G result in increased average cost savings (37\% and 18\% respectively) compared to NL-V2G and NL-V1G, but at the expense of charging compliance (83\% for L-V2G and 73\% for L-V1G). L-V2G and L-V1G increased cost savings come from estimating the nonlinear power ratings as constant, which causes mismatches between the control signal and the actual power rating capability at a given time step. This mismatch leads to EVs not achieving their session charging target, which results in lower power purchased from the grid and reduced charging target compliance. Additionally, Figure~\ref{Fig.CostCumulative} shows the cumulative costs of the V2G/V1G PF, NL-V2G, NL-V1G, and UC cases for the simulated year.

\subsection{V2G Energy Equivalent Mileage}
Most EVs have a battery warranty based on the production time and the drive mileage. To this end, V2G puts on additional discharges to the battery and may accelerate the expiration of the manufacturer warranty. In this section, we study how much additional energy is discharged in V2G and the equivalent mileage consumption to understand how much stress V2G would put on battery warranties.

Table~\ref{tab:energybalance} compares the total EVCS energy input (charged) and output (discharged) in NL-V2G and NL-V1G scenarios across all NYISO zones. On average, 7.3\% of the total energy charged in the NL-V2G scenario is used for discharging to arbitrage electricity prices. From these results, an equivalent mileage value for an EV participating in a V2G EVCS can be estimated by translating the energy output from the station to mileage through an EPA EV range estimate~\cite{epafuel}. This value becomes relevant when calculating the impact of V2G on EV warranty, which is regulated to be the first of 8 years or 100,000 mi (15 years or 150,000 mi proposed in California)~\cite{caligov}. As a case study, we use the Tesla Model X EPA estimated range of 348 mi for a full charge (100 kWh)~\cite{teslax}. Subsequently, a fraction of yearly energy output from the NYC EVCS (9.6 MWh), proportional to the number of charging sessions corresponding to a particular EV (139 sessions out of a total of 2967), is equivalent to a mileage value of 1565 mi. This corresponds to a 12\% increase in the average mileage driven per year (USDOT average is 13500 mi~\cite{DOT}), which would lead to passing 100,000 mi approximately nine months earlier than the baseline average mileage. Additionally, using the EVCS total energy output, an EPA estimated range of 348 mi and the cost savings achieved by V2G result in an incremental EVCS operating cost savings benefit of \$0.125/kWh and \$0.036/mi. Using our proposed method to estimate V2G equivalent mileage under different control policies would increase accuracy when performing long-term EVCS TEA studies.


\begin{table}[t]
\footnotesize
\caption{Energy Balances (in MWh) for a 1-year EVCS simulation}
\centering
\begin{tabular}{lcccc}
\hline
\hline
Zone & NYC & NORTH & WEST & LONGIL\\
\hline
NL-V1G Charged                   &152.44   &153.38   &153.40   &152.59\\
NL-V2G Charged              &163.85   &161.08   &177.48   &168.65\\
NL-V2G Discharged             &9.63   &5.84   &20.25   &13.84\\
\hline
\hline
\end{tabular}
\label{tab:energybalance}
\end{table}

\subsection{Computation Times}

Table \ref{tab:comptimes} shows computation times for 1-year simulations of three cases all using nonlinear battery models: 1) PF-MILP: nonlinear V2G optimization with perfect price forecast formulated using MILP and solved using Gurobi; 2) PF-DP: nonlinear V2G optimization with perfect price forecast solved with the proposed algorithm, note that in this case there is no uncertainty, so the proposed algorithm is essentially dynamic programming; 3) NL-V2G: nonlinear V2G optimization solved using the proposed stochastic dynamic programming algorithm. Hence, PF-MILP and PF-DP are deterministic, while SDP is stochastic.

The computation time result shows the computation tractability of the proposed algorithm in both deterministic and stochastic optimization. The comparison between PF-MILP and PF-DP is an apple-to-apple comparison as both algorithms solve a deterministic EVCS problem. PF-DP, the deterministic version of our proposed algorithm yields an average result within 1.5\% of the solution times provided by the MILP formulation, while the computation time is around 60x faster. 
Our proposed SDP-based algorithm achieves a computation time 7.5x faster than the MILP. Note that SDP is solving multi-stage stochastic optimization  while the MILP is solving deterministic optimization, while both use nonlinear battery models.  Therefore, our proposed algorithm can also be considered a faster, more efficient, and open-source alternative for solving deterministic smart charging control case studies.

\begin{table}[t]
\footnotesize
\caption{Computation times (in seconds) for a 1-year V2G simulation}
\centering
\begin{tabular}{lcccc}
\hline
\hline
Zone & NYC & NORTH & WEST & LONGIL\\
\hline
PF-MILP                   &5567   &6123   &6117   &6214\\
PF-DP              &97   &93   &90   &97\\
NL-V2G              &830   &825   &829   &814\\
\hline
\hline
\end{tabular}
\label{tab:comptimes}
\end{table}

\section{Conclusion}\label{sec:conc}

We proposed and tested an EVCS controller based on a nonlinear analytical stochastic dynamic programming algorithm and least-laxity first scheduling. Using historical prices from New York State, our proposed V2G algorithm achieved 24\% to 56\% of EVCS operating cost savings compared to uncontrolled charging while maintaining a 95\% charging target compliance and accounting for EV battery nonlinear behavior and price uncertainty in real-time. Our study covers smart charging in which EVs are not discharged to the grid. Still, our approach provides on average 17\% cost savings by responding to grid price variations compared to uncontrolled charging. We also show the importance of considering nonlinear battery models in V2G optimization, in which the battery power rating and efficiencies are dependent on the SoC, which is critical to ensure the EV meets its charging target while responding to time-varying prices. Finally, the proposed algorithm is open-source and not requiring any thrid party solvers, while the computation time surpasses commercial solvers. Hence, our approach is suitable for real-world implementations and scale-up for large-scale EV fleet management.

In the future, we plan to improve the proposed approach in several directions. The first is to integrate the solution method with data-driven probability price prediction methods. The current solution to the V2G problem using stochastic dynamic programming still requires a Markov process to be trained using historical price data. However, designing the Markov process can be complicated and limited by the quantity of historical price data. Second, our result still shows that V2G is more likely to miss charging targets due to the feeder capacity constraints, we will investigate approaches to manage the charging constraint better and improve charging compliance. Finally, we will test our proposed algorithm using more sophisticated charging scenarios such as using heterogeneous EV fleets and considering local renewable generation and study the connection between driving patterns and zone prices with the control policy performance.

\bibliographystyle{IEEEtran}	
\bibliography{IEEEabrv,literature_v1}		

\appendix

\subsection{Stochastic Price Arbitrage using Markov Process}

We discretize the stochastic real-time price as a first-order Markov Process with N nodes per time step ($\{\pi_{t,i}|\, t\in \mathcal{T},\, i\in N\}$), and a stage transition probability $\rho_{i,j,t}$ indicating the probability of transitioning from price node $i$ at time period $t$ to price node $j$ at time period $t+1$. Thus, we reformulate the stochastic arbitrage problem in \eqref{eq1} using discretized Markov price processes as 
\begin{subequations}\label{eq5}
\begin{align}
    \begin{split}
    Q_{t-1,i}(e_{t-1}) &= \max_{b_t, p_t} \pi_{t,i} \cdot  (p_t-b_t) - c p_t + V_{t,i}(e_t)
    \end{split} \\
    V_{t,i}(e_t) &= \sum_{j\in\mathcal{N}} \rho_{i,j,t}  \cdot Q_{t,j}(e_{t}) \label{eq:obj2}
\end{align}
subject to the EV constraints \eqref{p1_c1}--\eqref{p1_c4}.
\end{subequations}

\subsection{Value Function Computation}\label{app:vf}

Our proposed algorithm is based on the following result that updates $q_{t-1,i}$ from $v_{t,i}$, where $q_{t-1,i}$ is the derivative of $Q_{t-1,i}$ and $v_{t,i}$ is the derivative of $V_{t,i}$~\cite{xu2020vf}
\begin{align}\label{eq6}
    &q_{t-1,i}(e) = \\
    &\begin{cases}
    [1+\eta{\cdot}{\partial B}/{\partial e}+B{\cdot}{\partial \eta}/{\partial e}]{\cdot}v_{t,i}(e+B\eta)-\pi{\cdot}{\partial B}/{\partial e}  & \\ \indent\indent\text{if $\pi_{t,i}\leq v_{t,i}(e+B\eta){\cdot}\eta$} \\
    \pi_{t,i}{\cdot}[1/\eta+(B/\eta){\cdot}({\partial \eta}/{\partial e})]  & \\ \indent\indent \text{if $ v_{t,i}(e+B\eta){\cdot}\eta < \pi_{t,i} \leq v_{t,i}(e){\cdot}\eta$} \\
    v_{t,i}(e) & \\ \indent\indent \text{if $ v_{t,i}(e){\cdot}\eta < \pi_{t,i} \leq [v_{t,i}(e)/\eta + c]^+$} \\
    (\pi_{t,i}-c){\cdot}[\eta+(P/\eta){\cdot}({\partial \eta}/{\partial e})]-P{\cdot}{\partial c}/{\partial e} & \\ \indent\indent \text{if $ [v_{t,i}(e)/\eta + c]^+ < \pi_{t,i}$} \text{$ \leq [v_{t,i}(e-P/\eta)/\eta + c]^+$} \\
    [1-(1/\eta){\cdot}({\partial P}/{\partial e})+{P{\cdot}(1/\eta)}^2{\cdot}({\partial \eta}/{\partial e})]{\cdot}v_{t,i}(e-P/\eta)\\+(\pi_{t,i}-c){\cdot}{\partial P}/{\partial e}-P{\partial c}/{\partial e} & \\ \indent\indent \text{if $\pi_{t,i} > [v_{t,i}(e-P/\eta)/\eta + c]^+$}  \nonumber
    \end{cases}
\end{align}

Note that $\eta$, B, P and c depend on the starting SoC e as formulated in \eqref{eq1}. The function form of these nonlinear parameters in \eqref{eq6} is omitted to simplify the mathematical representation.

We restate the solution algorithm to compute the value function from~\cite{zjx2022} which enforces a final SoC value higher than a given threshold $e^{f}$. In this algorithm implementation we discretize $v_{t,i}$ into a set of $\mathcal{M}$ segments with value and SoC pairs 
\begin{align}
    \hat{v}_{t,i} = \{\nu_{t,i,m} | m\in \mathcal{M}\}
\end{align}
associated with equally spaced SoC segments $e_{t,m}$. In our implementation, we discretized the SoC into 1000 segments. The valuation algorithm is listed as following
\begin{enumerate}
    \item Set $T\to t$ to start from the last time period; initialize the final value-to-go function segments $\nu_{T,m}$ to zeros for $e_{T,m} > e^{f}$ and to a very high value (we use \$1000/MWh) for $e_{T,m} \leq e^{f}$. Note that the final value function does not depends on price nodes.
    \item Go to the earlier time step by setting $t-1 \to t$.
    \item During period $t$, go through each price node for $i\in\mathcal{N}$ and value function segment $m\in \mathcal{M}$. Update the charge and discharge efficiency corresponding to the SoC segment, compute \eqref{eq2} and store $q_{t-1,i}(e)$; note that here $q_{t-1,i}(e_{t-1,m})$ is also discretized with the same granularity as the value function.
    \item Calculate the value function of the previous time step as
    \begin{align}
        \nu_{t-1,i,m} = \textstyle \sum_{j\in\mathcal{N}} \rho_{i,j,t}  q_{t-1,j}(e_{t,m})
        \label{eq13}
    \end{align}
    which is the derivative of \eqref{eq:obj2}.
    \item Go to step 2) until reaching the first time step.
\end{enumerate}

\subsection{Control Policy for Single Storage Device}\label{app:cp}
We restate the control policy from~\cite{zjx2022}. After the value function computation step, control can be executed by responding to realized market prices and looking for the closest price node $\pi_{t,i}$ such that $\underline{\pi}_{t,i} \leq \lambda_t < \overline{\pi}_{t,i}$, then the storage control decision is updated as
\begin{subequations}
\begin{align}
    p_t &= \min\{\hat{p}_t, e_{t-1}\eta\} \label{eqa4:1}\\
    b_t &= \min\{\hat{b}_t, (E-e_{t-1})/\eta\} \label{eqa4:2}
\end{align}
where $\hat{p}_t$ and $\hat{b}_t$ are calculated as
\begin{align}
    &\{\hat{p}_t, \hat{b}_t\} = \nonumber\\
    &\begin{cases}
    \{0,B\} & \text{if $\lambda_t\leq v_{t,i}(e+B\eta)\eta$} \\
    \{0, \alpha\}  & \text{if $ v_{t,i}(e+B\eta)\eta < \lambda_t \leq v_{t,i}(e)\eta$} \\
    \{0,0\} & \text{if $ v_{t,i}(e)\eta < \lambda_t \leq [v_{t,i}(e)/\eta + c]^+$} \\
    \{\beta,0\} & \text{if $ [v_{t,i}(e)/\eta + c]^+ < \lambda_t$} \\
    & \quad\text{$ \leq [v_{t,i}(e-P/\eta)/\eta + c]^+$} \\
    \{P,0\} & \text{if $\lambda_t > [v_{t,i}(e-P/\eta)/\eta + c]^+$} 
    \end{cases} \label{eqa4:3}
\end{align}
in which $\alpha$ and $\beta$ are given as follows
\begin{align}
    \alpha &= (v^{-1}_{t,i}(\lambda_t/\eta)-e_{t-1})/\eta \label{eqa4:4}\\
    \beta &= (e_{t-1} - v^{-1}_{t,i}((\lambda_t-c)\eta))/\eta \label{eqa4:5}
\end{align}
where $v^{-1}_{t,i}$ is the inverse function of $v_{t,i}$.
\end{subequations}\label{eqa4}

\eqref{eqa4:1} and \eqref{eqa4:2} enforce the battery SoC constraints over the discharge  $\hat{p}_t$ and charge $\hat{b}_t$ decisions. \eqref{eqa4:3} calculates control decisions and following the same principle as to \eqref{eq1} but use the observed price $\lambda_t$ instead of the price nodes $\pi_{t,i}$. The control policy uses the approximated values of B, P, $\eta$ and c (all 4 could be nonlinear) corresponding to the current SoC.

\subsection{MILP Formulation}\label{app:milp}

We start by showing below the multi-period arbitrage formulation, which is equivalent to our proposed stochastic dynamic programming formulation if assuming a deterministic price process $\pi_t$ and also add a penalty function $C_{end}$ and a weighing parameter $\alpha$ to represent the cost associated with missing a final SoC target (the choice of $\alpha$ must achieve a comparable charging compliance target performance as the deterministic dynamic programming solution):
\begin{subequations}
\begin{gather}
\max_{p_t,b_t} \quad - \alpha C_{end}+\sum^T_t\pi_{t}{\cdot}(p_t-b_t) - cp_t
\textbf{ s.t.} \quad \text{(1c),(1e),(1f)}\nonumber
\end{gather}
\end{subequations}

We modify this model to a MILP model for the variable power rating, discharge penalty and efficiency benchmark calculation with ten SoC-efficiency segment pairs as
\begin{subequations}
\begin{gather}
\max_{p_{k,t},b_{k,t}} \quad - \alpha C_{end}+\sum^T_t\sum_k^K\pi_{t}{\cdot}(p_{k,t}-b_{k,t}) - c_{k}p_{k,t} \label{p2_obj}\\
\textbf{s.t.} \quad 0 \leq \sum_k^Kb_{k,t}/B_{k} \leq 1,\; 0\leq \sum_k^K p_{k,t}/P_{k} \leq 1 \label{p2_c1}\\
e_{k,t} - e_{k,t-1} = -p_{k,t}/\eta_k + b_{k,t}\eta_k \label{p2_c2}\\
E_1u_{1,t} \leq e_{1,t} \leq E_1 \label{p2_c3}\\
E_ku_{k,t} \leq e_{k,t} \leq E_ku_{k-1,t}, \quad \forall k\in\{2,...,K-1\} \label{p2_c4}\\
0 \leq e_{K,t} \leq E_Ku_{K-1,t}\label{p2_c5}\\
C_{end} = (\sum_k^K e_{k,T}-e^f)^2
\end{gather}
\end{subequations}
where $k$ is the index of the nonlinear power rating, efficiency and cycling penalty approximation segments (10 segments in this case). \eqref{p2_obj} is the objective function which sums up all segments. \eqref{p2_c1} and \eqref{p2_c2} are the power rating constraints and energy storage evolution constraints implement on all segments. \eqref{p2_c3}-\eqref{p2_c5} model the piece-wise linear approximation to the battery nonlinear parameter curves with binary variables $u_{k,t}$, which enforce the lower SoC segment must be full before upper SoC segments can take on non-zero values. For this paper's simulations, an $\alpha$ value of 1000000 is used. 

In the EVCS context, this optimization step fits in step 2) of the solution algorithm proposed in 4.3 (it replaces the value function calculation step). Every time an EV arrives at the EVCS, the MILP will provide an optimal schedule for the EV to follow for the remainder of the session. If the EV is deviated from the provided optimal schedule due to a low priority assigned during the LLF scheduling step, a new optimal schedule for the remainder of the session will be computed. 


%

\end{document}